\documentclass[pra,aps,showpacs,showkeys,twocolumn,twoside]{revtex4}
\usepackage{amsmath}
\usepackage{amssymb}
\usepackage{amscd}

\usepackage{latexsym}
\usepackage{epsfig}
\usepackage{graphics}

\newtheorem{defi}{Definition}
\newtheorem{lemma}[defi]{Lemma}
\newtheorem{thm}[defi]{Theorem}
\newtheorem{cor}[defi]{Corollary}
\newtheorem{rem}[defi]{Remark}
\newtheorem{prop}[defi]{Proposition}
\newtheorem{exempel}[defi]{Example}

\newcommand{\qed}{\hfill $\Box$}
\newcommand{\tr}{{\operatorname{Tr}}}
\newcommand{\id}{{\operatorname{id}}}

\newcommand{\bra}[1]{{\langle{#1}|}}
\newcommand{\ket}[1]{{|{#1}\rangle}}
\newcommand{\ketbra}[1]{{\ket{#1}\!\bra{#1}}}
\newcommand{\C}{{\mathbb{C}}}

\newcommand{\fset}[1]{{\mathcal{#1}}}

\newcommand{\1}{{\openone}}

\newlength{\blank}
\newlength{\equalsign}
\newenvironment{beweis}[1][{\hspace{-\blank}}]{{\noindent\emph{Proof~{#1}.\ }}}{\hfill $\Box$\vskip 0.5\baselineskip}
\newenvironment{expl}[1][{}]{\begin{exempel} {#1}\normalfont}{\end{exempel}}

\begin{document}

\title{Structure of states which satisfy strong subadditivity\protect\\
of quantum entropy with equality}
\author{Patrick Hayden}
\email{patrick@cs.caltech.edu}
\affiliation{Institute for Quantum Information, Caltech 107--81, Pasadena, CA 91125, USA}
\author{Richard Jozsa}
\email{richard@cs.bris.ac.uk}
\affiliation{Department of Computer Science, University of Bristol,\\
Merchant Venturers Building, Woodland Road, Bristol BS8 1UB, United Kingdom}
\author{D\'{e}nes Petz}
\email{petz@math.bme.hu}
\affiliation{Department for Mathematical Analysis, Mathematical Institute,\\
Budapest University of Technology and Economics,
Egry J\'{o}zsef utca 2, H--1111 Budapest, Hungary}
\author{Andreas Winter}
\email{winter@cs.bris.ac.uk}
\affiliation{Department of Computer Science, University of Bristol,\\
Merchant Venturers Building, Woodland Rd, Bristol BS8 1UB, United Kingdom}
\date{22nd August 2003}


\begin{abstract}
  We give an explicit characterisation of the quantum states which saturate the
  strong subadditivity inequality for the von Neumann entropy.
  By combining a result of Petz characterising the equality case for the 
  monotonicity of relative entropy with a recent theorem by Koashi and Imoto,
  we show that such states will have the form of a so--called
  short quantum Markov chain, which in turn implies that two of the systems 
  are independent conditioned on the third, in a physically meaningful sense.
  This characterisation simultaneously generalises known necessary and
  sufficient entropic conditions for quantum error correction as well as
  the conditions for the achievability of the Holevo bound on accessible
  information.
\end{abstract}



\maketitle

\section{Introduction}
\label{sec:introduction}
The von Neumann entropy~\cite{von:neumann:entropy}
$$S(\rho)=-\tr\rho\log\rho,$$
of a density operator 
$\rho$ on a finite dimensional Hilbert space ${\cal H}$
shares many properties with its classical counterpart,
the Shannon entropy
$$H(P)=-\sum_{x\in\fset{X}} P(x)\log P(x)$$
of a probability distribution $P$ on a discrete set $\fset{X}$.
(All logarithms in this work are understood to be to base $2$. Also, we will use
the terms ``state'' and ``density operator'' interchangeably.)
For example, both are nonnegative, and equal to $0$ if and only if the
state (distribution) is an extreme point in the set of all
states (distributions), i.e. if $\rho$ is pure ($P$ is a point mass).
Both are concave and, moreover, both are subadditive: for a state
$\rho_{AB}$ on a composite system ${\cal H}_A\otimes{\cal H}_B$
with reduced states
$$\rho_A=\tr_B\bigl(\rho_{AB}\bigr),\quad \rho_B=\tr_A\bigl(\rho_{AB}\bigr),$$
it holds that
$$S(\rho_{AB}) \leq S(\rho_A)+S(\rho_B).$$
A directly analogous inequality holds for a distribution over a 
product set and its marginals.
(Many more properties of $S$ are collected in the review by
Wehrl~\cite{wehrl} and in the monograph \cite{ohya:petz}.)
\par
We shall view von Neumann entropy as a generalisation of Shannon
entropy~\cite{shannon:info}
in the following precise way: if the set $\fset{X}$ labels
an orthonormal basis $\bigl\{\ket{x}:x\in\fset{X}\bigr\}$ of ${\cal H}_X$
we can construct the state
$$\rho_P=\sum_x P(x)\ketbra{x}$$
corresponding to the distribution $P$. This clearly defines an affine linear
map from distributions into states.
It is then straightforward to check that
$$S(\rho_P)=H(P),$$
so all properties of von Neumann entropy of a single system also
hold for Shannon entropy of a single distribution.
\par
Similarly, for a distribution $P$ on a cartesian product $\fset{X}\times\fset{Y}$,
we use the tensor product basis
$$\bigl\{\ket{xy}=\ket{x}\otimes\ket{y}:x\in\fset{X},y\in\fset{Y}\bigr\}$$
to define the state $\rho_P$ on ${\cal H}_X\otimes{\cal H}_Y$.
Again, it is straightforward to check that reduced states correspond to
taking marginals:
$$\tr_Y\bigl(\rho_P\bigr) = \rho_{P|_{\fset{X}}},\quad
  \tr_X\bigl(\rho_P\bigr) = \rho_{P|_{\fset{Y}}}.$$
Hence all entropy relations for bipartite states also hold
for bipartite distributions.
\par
In~\cite{lieb:ruskai:SSA} Lieb and Ruskai proved the remarkable relation
\begin{equation}
  \label{eq:SSA}
  S(\rho_{AB})+S(\rho_{BC}) \geq S(\rho_{ABC})+S(\rho_B),
\end{equation}
with a tripartite state $\rho_{ABC}$ on the system
${\cal H}_A\otimes{\cal H}_B\otimes{\cal H}_C$. It clearly generalises
the previous subadditivity relation, which is recovered for a
trivial system $B$: ${\cal H}_B=\C$. In fact, this inequality plays a 
crucial role in nearly every nontrivial
insight in quantum information theory, from the famous
Holevo bound~\cite{holevo:bound} and the properties of the coherent
information~\cite{schumacher:et-al,schumacher:qec} to the recently proved additivity
of capacity for entanglement--breaking channels~\cite{shor:ent-break}.
\par
The present investigation aims to resolve the problem of characterising
the states which satisfy this relation with equality: the main result
is theorem~\ref{thm:SSA:eq}.
Roughly speaking, the strong subadditivity inequality expresses the fact
that discarding a subsystem of a quantum system is a dissipative operation,
in the sense that it can only destroy correlations with the rest of the world. 
Our work, therefore, can be interpreted as providing a detailed description 
of the conditions under which the act of discarding a quantum system can be 
locally reversed on a particular input.
We restrict ourselves to finite dimensional systems in
this paper. The question of whether a similar result holds in infinite 
dimension is left open.
\par
The rest of the paper is organised as follows.
In section~\ref{sec:classical} we will review the
case of probability distributions: there the solution to our problem
is easy to obtain, and in fact well--known. This will provide the
intuitive basis for understanding our main result.
After that, in section~\ref{sec:rel:ent} we review quantum relative
entropy and the relation of its monotonicity property
to the strong subadditivity inequality.
Section~\ref{sec:petz} presents a condition given by Petz for 
equality in the monotonicity of relative entropy, while
section~\ref{sec:equality} presents and proves our main result,
a structure theorem for states which satisfy strong subadditivity
with equality. An essential step is the application of a recent
result of Koashi and Imoto~\cite{koashi:imoto} for which we give a
short but non--constructive algebraic proof in the appendix.
In section~\ref{sec:applications}, we show how the entropic conditions
for quantum error correction as well as the conditions for saturation
in the Holevo bound follow as easy corollaries from our structure theorem.
\par
The question of characterising the equality case of strong subadditivity
as well as of the monotonicity of relative entropy was considered in
earlier work by Petz~\cite{petz:sufficient},
where it was related to the existence of quantum operations with
certain properties. Ruskai~\cite{ruskai:SSA:eq} has given a 
characterisation in terms of an operator equality, which can be
used to show that the states described in our main theorem~\ref{thm:SSA:eq}
are indeed equality cases (as she has informed us, this was pointed out to
her by M.~A.~Nielsen after~\cite{ruskai:SSA:eq} appeared).
Neither of these results is as explicit as one could wish for, however, because
while both give algebraic criteria which one can check on any given state,
they do not yield a simple description of all the states that satisfy equality.
This simple description is exactly what our theorem~\ref{thm:SSA:eq} provides.

\section{The classical case}
\label{sec:classical}
Let us first look at the classical case of probability distributions
and their Shannon entropies. The exposition is most conveniently
phrased in terms of random variables denoted $A,B,C$, taking
values in ${\cal A,B,C}$, respectively, with a joint
distribution
$$P_{ABC}(a,b,c)=\Pr\{A=a,B=b,C=c\}.$$
The distribution of $A$ is the marginal $P_A=P_{ABC}|_{\cal A}$
of the joint distribution to ${\cal A}$ and similarly for the other variables.
\par
Shannon~\cite{shannon:info} defined the \emph{mutual information}
$$I(A:B)=H(A)+H(B)-H(AB),$$
with $H(A)=H(P_A)$ and so on. It is not hard to show
that $I(A:B)\geq 0$ with equality if and only
if $A$ and $B$ are independent.
\par
\emph{Conditional mutual information} is defined as
\begin{equation*}
  I(A:C|B)=\sum_{b\in{\cal B}} P_B(b) I(A:C|B=b),
\end{equation*}
where $I(A:C|B=b)$ is the mutual information between the
variables $A$ and $C$ conditional on the event ``$B=b$'',
i.e.~$I(A|_{B=b}:C|_{B=b})$, with
\begin{equation*}\begin{split}
  \Pr\bigl\{ A|_{B=b}=a, C|_{B=b}=c \bigr\} &= \frac{P_{ABC}(a,b,c)}{P_B(b)} \\
                                            &=:P_{AC|B}(a,c|b).
\end{split}\end{equation*}
It is straightforward to check that with these definitions one has the
\emph{chain rule}
\begin{equation}
  \label{eq:chainrule}
  I(A:BC) = I(A:B) + I(A:C|B).
\end{equation}
This implies the formula
$$I(A:C|B) = H(AB)+H(BC)-H(ABC)-H(B).$$
Because the left hand side is by definition a convex combination
of mutual informations, each of which is always nonnegative, we obtain strong
subadditivity for classical distributions.
\begin{thm}
  \label{thm:SSA-eq:classical}
  $I(A:C|B)=0$ if and only if $A$ and $C$ are conditionally 
  independent given $B$, meaning
  $$\forall b\text{ s.t. }P_B(b)\neq 0\quad A|_{B=b},C|_{B=b}\text{ are independent.}$$
  This is the case if and only if
  \begin{equation}\begin{split}
    \label{eq:markov:classical}
    P_{ABC}(a,b,c) &= P_B(b)P_{A|B}(a|b)P_{C|B}(c|b) \\
                   &= P_A(a)P_{B|A}(b|a)P_{C|B}(c|b),
  \end{split}\end{equation}
  i.e.~iff $A$---$B$---$C$ is a Markov chain in this order.
\end{thm}
\begin{beweis}
  Clearly, the conditions are sufficient. Assume conversely that
  $I(A:C|B)=0$. By definition of the latter quantity, this implies
  that for all $b$ with $P_B(b)\neq 0$, $I(A:C|B=b)=0$.
  But this implies independence of $A|_{B=b}$ and $C|_{B=b}$.
  Hence, eq.~(\ref{eq:markov:classical}) follows:
  \begin{equation*}\begin{split}
    P_{ABC}(a,b,c) &= P_B(b)P_{AC|B}(a,c|b)          \\
                   &= P_B(b)P_{A|B}(a|b)P_{C|B}(c|b) \\
                   &= P_A(a)P_{B|A}(b|a)P_{C|B}(c|b).
  \end{split}\end{equation*}
\end{beweis}
\par
The remainder of the paper is devoted to describing the
quantum mechanical generalisation of this
equivalence between zero conditional mutual information, conditional 
independence,
the Markov property and the factorization of the joint distribution given
in eq.~(\ref{eq:markov:classical}).

\section{Relative entropy}
\label{sec:rel:ent}
Our approach to saturation of the strong subaddivity inequality will
be via the quantum relative entropy;
this quantity was defined by Umegaki~\cite{umegaki} for two
quantum states $\rho$ and $\sigma$ as
$$S(\rho\|\sigma)=\tr\bigl(\rho(\log\rho-\log\sigma)\bigr)$$
if the support of $\rho$ is contained in the support of $\sigma$,
and $+\infty$ otherwise. We note that this definition generalises
the familiar Kullback--Leibler divergence~\cite{kullback:leibler}
of two probability distributions,
just as von Neumann entropy generalises Shannon entropy.
\par
For a bipartite state $\rho_{AB}$ it is straightforward to check that
\begin{equation}
  \label{eq:mutualinfo:RE}
  S(\rho_{AB}\|\rho_A\otimes\rho_B) = S(\rho_A)+S(\rho_B)-S(\rho_{AB}),
\end{equation}
and the latter quantity is abbreviated $I(A:B)$, in formal extension
of the definition of Shannon's mutual information~\cite{shannon:info} to
quantum states.
\begin{expl}
  \label{expl:holevo}
  Let $\{p(x),\rho_x\}$ be an ensemble of quantum states
  on ${\cal H}$. The \emph{Holevo quantity $\chi$} is defined as
  $$\chi\bigl(\{p(x),\rho_x\}\bigr)
         = S\left(\sum_x p(x)\rho_x\right)-\sum_x p(x) S(\rho_x).$$
  Holevo~\cite{holevo:bound} showed that this quantity is an upper bound
  to the mutual information between $x$ and the outcomes $y$ of any
  particular measurement performed on the states $\rho_x$.
  \par
  It is easily seen that
  $$\chi\bigl(\{p(x),\rho_x\}\bigr) = I(A:B),$$
  with the bipartite state
  $$\rho_{AB}=\sum_x p(x)\ketbra{x}_A\otimes(\rho_x)_B.$$
\end{expl}
For a tripartite state $\rho_{ABC}$ we can also consider the information
$I(A:BC)$, which can be written as
\begin{equation}
  \label{eq:mutualinfo:RE:2}
  S(\rho_{ABC}\|\rho_A\otimes\rho_{BC}) = S(\rho_A)+S(\rho_{BC})-S(\rho_{ABC}).
\end{equation}
The difference between eqs.~(\ref{eq:mutualinfo:RE:2})
and~(\ref{eq:mutualinfo:RE}),  which by virtue of the classical
chain rule eq.~(\ref{eq:chainrule}) we might call the quantum conditional
mutual information $I(A:C|B)$ is, therefore,
\begin{equation}\begin{split}
  \label{eq:RE-RE}
  S(\rho_{ABC}\|\rho_A\otimes\rho_{BC})
                          &- S(\rho_{AB}\|\rho_A\otimes\rho_B)               \\
                          &\!\!\!\!\!\!\!\!\!\!\!\!\!\!\!\!\!\!\!\!\!\!\!\!\!
                           = S(\rho_{AB})+S(\rho_{BC})-S(\rho_{ABC})-S(\rho_B).
\end{split}\end{equation}
The right hand side here is nonnegative by strong subadditivity.
(Note that this is an important theorem in the quantum case despite being
an almost trivial observation classically.) The left hand
side, however, can be rewritten as $S(\rho\|\sigma)-S(T\rho\|T\sigma)$,
with the states $\rho=\rho_{ABC}$ and $\sigma=\rho_A\otimes\rho_{BC}$,
and the quantum operation $T=\tr_C$, the partial trace over ${\cal H}_C$,
which as a linear map can be written as $T=\id_{AB}\otimes\tr$.
\par
Now a theorem of Uhlmann~\cite{uhlmann} (proved earlier by
Lindblad~\cite{lindblad:mono}
for the finite--dimensional case of interest here)
says that for all states $\rho$ and
$\sigma$ on a space ${\cal H}$, and all quantum operations
$T:{\cal B}({\cal H})\rightarrow{\cal B}({\cal K})$,
\begin{equation}
  \label{eq:RE:mono}
  S(\rho\|\sigma) \geq S(T\rho\|T\sigma),
\end{equation}
so Uhlmann's theorem implies strong subadditivity and we have equality in the
latter if and only if there is equality in the former.

\section{The equality condition\protect\\ for relative entropy}
\label{sec:petz}
The formulation in the previous section
of strong subadditivity as a relative entropy
monotonicity under a partial trace operation
transforms the question for the equality conditions for the
former into the same question for the latter.
Note that by the very monotonicity relation, there is a ``trivial''
case of equality in eq.~(\ref{eq:RE:mono}), 
namely if there exists a quantum operation
$\widehat{T}$ mapping $T\rho$ to $\rho$ and $T\sigma$ to $\sigma$.
In fact, this is the only case of equality:
\begin{thm}[Petz~\cite{petz:sufficient}]
  \label{thm:petz}
  For states $\rho$ and $\sigma$, 
  $$S(\rho\|\sigma) = S(T\rho\|T\sigma)$$
  if and only if there exists a quantum operation $\widehat{T}$
  such that
  $$\widehat{T}T\rho=\rho,\quad \widehat{T}T\sigma=\sigma.$$
  Furthermore, on the support of $T\sigma$, $\widehat{T}$ can be
  given explicitly by the formula
  \begin{equation}
    \label{eq:transpose:channel}
    \widehat{T}\alpha=\sigma^{\frac{1}{2}}
                        T^*\left(
                             (T\sigma)^{-\frac{1}{2}}\alpha (T\sigma)^{-\frac{1}{2}}
                           \right)\sigma^{\frac{1}{2}},
  \end{equation}
  with the adjoint map $T^*$ of $T$:
  $$T^*(X)=\sum_i A_i^* X A_i, \text{  if  }
       T(\alpha)=\sum_i A_i \alpha A_i^*.$$
  \qed
\end{thm}
Observe that the definition of $\widehat{T}$ in 
eq.~(\ref{eq:transpose:channel}) depends on $\sigma$, thereby 
automatically ensuring that $\widehat{T}T\sigma=\sigma$. Sometimes, we
add the subscript $\sigma$ to $\widehat{T}$ to emphasize the dependence.
\begin{expl}
  \label{expl:holevo:bound}
  As in example~\ref{expl:holevo}, let $\{p(x),\rho_x\}$ be an ensemble
  of states on ${\cal H}$
  and define
  $$\rho_{AB}=\sum_x p(x)\ketbra{x}_A\otimes(\rho_x)_B.$$
  There we observed that, with $\sigma=\rho_A\otimes\rho_B$,
  $$\chi\bigl(\{p(x),\rho_x\}\bigr)=S(\rho\|\sigma).$$
  Now, let $\varphi$ be a quantum operation on ${\cal H}$
  (which could be a measurement), and form $T=\id_A\otimes\varphi_B$.
  Then
  \begin{equation}\begin{split}
    \label{eq:holevo:bound}
    \chi\bigl(\{p(x),\rho_x\}\bigr) &=    S(\rho_{AB}\|\rho_A\otimes\rho_B)    \\
                         &\geq S\bigl(T\rho_{AB}\|T(\rho_A\otimes\rho_B)\bigr) \\
                         &=    \chi\bigl(\{p(x),\varphi\rho_x\}\bigr),
  \end{split}\end{equation}
  which is (a generalistion of) the famous Holevo bound~\cite{holevo:bound}
  in the form of a data processing relation.
  \par
  Equality holds, according to theorem~\ref{thm:petz}, if and only if
  $\widehat{T}$ of eq.~(\ref{eq:transpose:channel}) maps $T\rho$ to $\rho$.
  Note that we may assume without loss of generality
  that $\sigma=\rho_A\otimes\rho_B$ is
  strictly positive. But it is straightforward to check that
  $$\widehat{T}_\sigma=\widehat{\id}_{\rho_A}\otimes\widehat{\varphi}_{\rho_B}
                      =\id\otimes\widehat{\varphi},$$
  hence we have equality in eq.~(\ref{eq:holevo:bound}) if
  and only if for all $x$, $\widehat{\varphi}\varphi\rho_x=\rho_x$.
\end{expl}
\begin{rem}
  \label{rem:barnum-knill}
  In~\cite{barnum:knill} the ``transpose channel'' $\widehat{T}$ of
  eq.~(\ref{eq:transpose:channel}), as it is called in~\cite{ohya:petz},
  makes an appearance in a slightly different context:
  there a set of states is subjected to a quantum channel, and
  the problem is to find the best recovery map which maximises a
  fidelity criterion for the original states and the images of the
  channel output states. It was shown that the error using
  $\widehat{T}$ is always at most twice as the minimum error under the
  optimal recovery map.
\end{rem}

\section{Structure of states with equality}
\label{sec:equality}
Let $\rho_{ABC}$ be a state on ${\cal H}_A\otimes{\cal H}_B\otimes{\cal H}_C$.
As we observed earlier, Uhlmann's theorem specialized to
the states $\rho_{ABC}$ and $\sigma_{ABC}=\rho_A\otimes\rho_{BC}$,
along with the map $T=\tr_C$, states that
$$S(\rho_{ABC}\|\rho_A\otimes\rho_{BC})
      \geq S(\rho_{AB}\|\rho_A\otimes\rho_B).$$
Consequently, theorem \ref{thm:petz} provides the condition for equality here:
$\widehat{T}T\rho=\rho$. Now, because $T=\id_A\otimes R_{BC}$, with
the restriction map $R_{BC}=\id_B\otimes\tr_C$,
and $\sigma$ is a tensor product, we obtain (compare
example~\ref{expl:holevo:bound}):
\begin{equation}
  \label{eq:form:of:hat-T}
  \widehat{T}=\id_A\otimes\widehat{R},
\end{equation}
with $\widehat{R}=\widehat{R}_{\rho_{BC}}$.
\par
Summarising, in the above monotonicity and hence in strong subadditivity
we have equality if and only if
\begin{equation}
  \label{eq:markov:quantum}
  \rho_{ABC} = (\id\otimes\widehat{R})\rho_{AB}.
\end{equation}
\par
We are now in a position to prove our main result:
\begin{thm}
  \label{thm:SSA:eq}
  A state $\rho_{ABC}$ on ${\cal H}_A\otimes{\cal H}_B\otimes{\cal H}_C$
  satisfies strong subadditivity (eq.~(\ref{eq:SSA})) with equality if and only
  if there is a decomposition of system $B$ as
  \begin{equation*}
    {\cal H}_B=\bigoplus_j {\cal H}_{b^L_j}\otimes{\cal H}_{b^R_j}
  \end{equation*}
  into a direct sum of tensor products, such that
  \begin{equation*}
    \rho_{ABC}=\bigoplus_j q_j \rho_{Ab^L_j}\otimes\rho_{b^R_jC},
  \end{equation*}
  with states $\rho_{Ab^L_j}$ on ${\cal H}_A\otimes{\cal H}_{b^L_j}$
  and $\rho_{b^R_jC}$ on ${\cal H}_{b^R_j}\otimes{\cal H}_C$,
  and a probability distribution $\{q_j\}$.
\end{thm}
\begin{beweis}
  The sufficiency of the condition is immediate. The proof of necessity will
  come from analysing the quantum Markov chain condition, 
  eq.~(\ref{eq:markov:quantum}).
  \par
  After defining the quantum operation $\varphi=\tr_C\circ\widehat{R}$,
  the Markov condition gives us
  \begin{equation}
    \label{eq:varphi:invariance}
    (\id\otimes\varphi)\rho_{AB}=\rho_{AB}.
  \end{equation}
  Consider an operator $M$ on ${\cal H}_A$ with $0\leq M\leq \1$, and
  define a state $\mu$ by
  $$p\mu = \tr_A\bigl(\rho_{AB}(M\otimes\1)\bigr),\ 
       p = \tr\bigl(\rho_{AB}(M\otimes\1)\bigr).$$
  Then, if $p\neq 0$, eq.~(\ref{eq:varphi:invariance}) implies that
  $\varphi(\mu)=\mu$. Varying the operator $M$ we obtain a family
  ${\bf M}$ of states on ${\cal H}_A$ invariant under $\varphi$.
  \par
  To this we can apply theorem~\ref{thm:koashi} from the
  appendix. We obtain a decomposition 
  \begin{equation}
    \label{eq:directsum:B}
    {\cal H}_B = \bigoplus_j {\cal H}_{b^L_j}\otimes{\cal H}_{b^R_j},
  \end{equation}
  such that every $\mu\in{\bf M}$ can be written
  $$\mu = \bigoplus_j q_j(\mu) \rho_j(\mu)\otimes\omega_j,$$
  with states $\rho_j(\mu)$ on ${\cal H}_{b^L_j}$ and
  $\omega_j$ on ${\cal H}_{b^R_j}$.
  This, in turn, easily implies the following structure for $\rho_{AB}$:
  \begin{equation}
    \label{eq:rho-AB}
    \rho_{AB} = \bigoplus_j q_j \rho_{Ab^L_j}\otimes\omega_{b^R_j}.
  \end{equation}
  To see this, introduce the quantum operation
  $$P_0(\xi)=\bigoplus_j \tr_{b^R_j}\bigl(\Pi_j\xi\Pi_j\bigr)\otimes\omega_j,$$
  on ${\cal H}_B$, where $\Pi_j$ is the orthogonal projector
  onto the subspace ${\cal H}_{b^L_j}\otimes{\cal H}_{b^R_j}$ in
  eq.~(\ref{eq:directsum:B}). (Its dual $P_0^*$ is the subalgebra projection
  from the appendix, where it is denoted the same way.) 
  Then it is easy to calculate, for
  arbitrary operators $M$ and $N$ bounded between $0$ and $\1$:
  \begin{equation*}\begin{split}
    \tr\bigl( \rho_{AB}(M\otimes N) \bigr)
               &= p\tr(\mu N)                                                \\
               &= p\tr\bigl( P_0(\mu)N \bigr)                                \\
               &= p\tr\bigl( \mu P_0^*(N) \bigr)                             \\
               &=  \tr\Bigl( \rho_{AB}\bigl(M\otimes P_0^*(N)\bigr) \Bigr)             \\
               &= \tr\Bigl( \bigl((\id\otimes P_0)\rho_{AB}\bigr)(M\otimes N) \bigr). \\
  \end{split}\end{equation*}
  By linearity, this holds for all operators in place of $M\otimes N$, so
  $$\rho_{AB} = (\id\otimes P_0)\rho_{AB},$$
  implying eq.~(\ref{eq:rho-AB}), because
  $$(\id\otimes P_0)\xi
    = \bigoplus_j \tr_{b^R_j}\bigl( (\1\otimes\Pi_j)\xi(\1\otimes\Pi_j^*)\bigr)
                                                                 \otimes\omega_j.$$
  \par
  But theorem~\ref{thm:koashi} also gives information about $\varphi$:
  introduce an environment ${\cal H}_E$ in state $\varepsilon$
  and a unitary $U$ on ${\cal H}_B\otimes{\cal H}_C\otimes{\cal H}_E$
  such that
  $$\widehat{R}(\alpha)
      =\tr_E\bigl(U(\alpha\otimes\ketbra{0}\otimes\varepsilon)U^*\bigr),$$
  with a standard state $\ket{0}\in{\cal H}_C$.
  Because a further trace over $C$ gives us $\varphi$, we obtain
  the following form for $U$ (with ${\cal E}={\cal H}_C\otimes{\cal H}_E$):
  \begin{equation}
    \label{eq:form:of:U}
    U=\bigoplus_j \1_{{\cal H}_{b^L_j}}\otimes U_j,
  \end{equation}
  with $U_j$ a unitary on ${\cal H}_{b^R_j}\otimes{\cal E}$.
  \par
  Putting eqs.~(\ref{eq:rho-AB}) and~(\ref{eq:form:of:U})
  together, we finally get:
  \begin{equation*}\begin{split}
    \rho_{ABC} &= (\id_A\otimes\widehat{R})\rho_{AB}                         \\
               &= \tr_E\bigl((\1_A\otimes U)\rho_{AB}(\1_A\otimes U^*)\bigr) \\
               &= \bigoplus_j q_j \rho_{Ab^L_j}\otimes
                               \tr_E\bigl(U(\omega_{b^R_j}\otimes\ketbra{0})U^*\bigr) \\
               &= \bigoplus_j q_j \rho_{Ab^L_j}\otimes\rho_{b^R_jC},
  \end{split}\end{equation*}
  which is what we wanted to prove.
\end{beweis}
\par
Quantum Markov states on the infinite tensor product of matrix algebras
$\bigotimes_{i=-\infty}^\infty M_n({\C})^{(i)}$ were introduced by Accardi and
Frigerio~\cite{AF}. Let ${\cal A}_m$ be the subproduct of the factors with 
superscript $i \leq m$. Then ${\cal A}_m \subset {\cal A}_{m+1}$. A state $\rho$
of the infinite tensorproduct is called {\it Markovian} if for every
integer $m$ there exists a unital completely positive mapping
$T_{m,m+1} : {\cal A}_{m+1} \rightarrow {\cal A}_{m}$ which
leaves the state $\rho$ (restricted to ${\cal A}_m$)
invariant and the subalgebra ${\cal A}_{m-1}$ fixed.
Accardi and Frigerio call the mapping ${\cal E}_{m,m+1}$
{\it quasi--conditional expectation}; its dual is the quantum analogue
of the Markov kernel in classical probability theory.
\par
Assume that ${\cal A}_{m-1}={\cal B}({\cal H}_A)$,
$M_n({\C})^{(m)}={\cal B}({\cal H}_B)$ and
$M_n({\C})^{(m+1)}= {\cal B}({\cal H}_C)$. If the equality in
strong subadditivity is satisfied in this setting, then we have
eq.~(\ref{eq:form:of:hat-T}) and the dual of $\widehat{T}$ is a
quasi--conditional expectation. Therefore the equality in strong 
subadditivity for every $m$ yields a quantum Markov state on
the infinite system. This property characterises quantum Markov states,
see e.g.~\cite{ohya:petz}, p. 201.
\par
We propose to call a state as in eq.~(\ref{eq:markov:quantum})
a \emph{short quantum Markov chain} (as opposed to the infinite
chains introduced in~\cite{AF}), since we require the existence
of the quasi--conditional expectation only for
${\cal B}({\cal H}_A \otimes {\cal H}_B \otimes {\cal H}_C) \rightarrow
{\cal B}({\cal H}_A \otimes {\cal H}_B)$;
note that the analogous quasi--conditional expectation
${\cal B}({\cal H}_A \otimes {\cal H}_B) \rightarrow {\cal B}({\cal H}_A)$
exists trivially because the subalgebra to be left invariant is $\C$.
\par
\begin{cor}
  \label{cor:separable}
  For a state $\rho_{ABC}$ satisfying strong subadditivity with equality:
  $$I(A:C|B) = S(\rho_{AB})+S(\rho_{BC})-S(\rho_{ABC})-S(\rho_B) = 0,$$
  the marginal state $\rho_{AC}$ is separable.
  \par
  Conversely, for each separable state $\rho_{AC}$ there exists an
  extension $\rho_{ABC}$ such that $I(A:C|B)=0$.
  \qed
\end{cor}
Tucci~\cite{tucci} has given a criterion for separability based on
quantum conditional mutual information. Our above result shows that
in his Theorem 1, only conditions 1 and 2 are needed, while 3 and 4
are redundant.

\section{Applications}
\label{sec:applications}
Theorem \ref{thm:SSA:eq} provides a convenient framework for synthesizing
many previously known facts in quantum information theory. To illustrate
the method of its application, we present a couple of special
cases from the literature.
\begin{expl}
  \label{expl:qec}
  The fundamental problem in quantum error correction is to determine 
  when the effect of a quantum operation $\varphi$ acting on half of a 
  pure entangled state can be perfectly reversed. 
  Define the coherent information
  $$I_c(\sigma,\varphi) = S(\varphi\sigma) - 
                            S\bigl( (\id_A \otimes \varphi)\Phi_\sigma \bigr),$$
  where $\Phi_\sigma$ is any purification of $\sigma$ to system $A$.
  In~\cite{schumacher:qec} it was shown that there exists a quantum operation
  $\hat{\varphi}$ such that
  $$(\id_A \otimes \hat{\varphi}\varphi)\Phi_\sigma = \Phi_\sigma$$
  if and only if 
  \begin{equation}
    \label{eqn:coherent:EQ}
    I_c(\sigma,\varphi) = S(\sigma).
  \end{equation}
  By the Stinespring dilatation theorem~\cite{stinespring}, we may assume that
  $$\varphi\sigma = \tr_C \bigl( U_{BC}(\sigma\otimes\psi)U_{BC}^* \bigr)$$
  for a unitary operator $U_{BC}$ and pure state ancilla $\psi$ on system $C$.
  If we let
  $\ket{\omega} = (\1_A \otimes U_{BC})(\ket{\Phi_\sigma} 
                      \otimes \ket{\psi})$
  then, taking mutual informations with
  respect to the state $\omega=\ketbra{\omega}$,
  \begin{align*}
    S(\sigma)           &= I(A:BC) - S(A)\text{ and} \\
    I_c(\sigma,\varphi) &= I(A:B)-S(A).
  \end{align*}
  Therefore, eq.~(\ref{eqn:coherent:EQ})
  holds iff $I(A:BC)=I(A:B)$. By theorem~\ref{thm:SSA:eq}, we can conclude that
  $$\omega = \bigoplus_j q_j \omega_{Ab_j^L} \otimes \omega_{b_j^R C}.$$
  The recovery procedure $\hat{\varphi}$ given the state
  $$(\id_A \otimes \varphi)\Phi_\sigma = \tr_C \omega$$
  is then obvious: first measure $j$ before preparing the state 
  $\omega_{b_j^R C}$ on ${\cal H}_{b_j^R} \otimes {\cal H}_C$. Next, perform
  $U_{BC}^*$ and discard the fixed ancilla state $\psi$. The output 
  is exactly $\Phi_\sigma$. 
  As an aside, the reason that the solution to this
  problem was accessible without the results of the present paper is that 
  a closer examination of the situation reveals that, because $\Phi_\sigma$
  is pure, only the equality conditions 
  for the usual subadditivity inequality are required in the construction
  of the reversal map. Strong subadditivity, in this case, is superfluous.
\end{expl}
\par
\begin{expl}
  \label{expl:holevo:bound:eq}
  Returning to our investigation of the Holevo bound from
  example~\ref{expl:holevo:bound}, let $\{p(x),\rho_x\}$ be an ensemble of states
  on ${\cal H}$, $\rho_{AB} = \sum_x p(x)\ketbra{x}_A \otimes (\rho_x)_B$
  and $\varphi$ a quantum operation on the $B$ system. 
  Again by the Stinespring dilation theorem, after possibly adjoining a fixed ancilla
  and performing a common unitary operation to all the $\rho_x$, we may
  assume without loss of generality that 
  $B = \tilde{B} \tilde{C}$ and $\varphi = \tr_{\tilde{C}}$. 
  An application of
  theorem~\ref{thm:SSA:eq} then gives the conditions under which 
  $$\chi(\{p(x),\rho_x\})=\chi(\{p(x),\varphi\rho_x\}).$$
  Namely,
  $$\rho_{A\tilde{B}\tilde{C}} = 
        \bigoplus_j q_j \sum_x p(x|j) \ketbra{x}_A \otimes (\rho_x)_{\tilde{b}_j^L} 
                                           \otimes \omega_{\tilde{b}_j^R \tilde{C}},$$
  where $p(x|j)q_j$ is the joint probability distribution for $(x,j)$ 
  and $\omega_{\tilde{b}_j^R \tilde{C}}$ does not depend on $x$.
  In the special case where $\varphi$ corresponds to a measurement operation,
  the additional constraint $[\varphi \rho_x, \varphi \rho_{x'}] = 0$ must
  hold because the output system is classical. 
  Given the form of $\rho_{A\tilde{B}\tilde{C}}$, this implies
  $[(\rho_x)_{\tilde{b}_j^L},(\rho_{x'})_{\tilde{b}_j^L}] = 0$ and, in turn,
  that the states $\{\rho_x\}$ all commute. In the language of quantum
  information, we have found that the accessible information of an
  ensemble is equal to its Holevo quantity if and only if all the states
  in the ensemble commute. This condition for equality actually appeared
  in Holevo's original paper~\cite{holevo:bound}.
\end{expl}

\section{Discussion}
\label{sec:discussion}
We have exhibited the explicit structure of the tripartite states $\rho_{ABC}$
which satisfy strong subadditivity with equality. Not only are they short 
quantum Markov chains in the sense of~\cite{AF},
it is even the case that the $A$ and $C$ systems are conditionally
independent given $B$, in a physically meaningful sense:
there is information in the $B$ system which can be obtained by
a non--demolition measurement, conditioned upon which the quantum state
factorises.
\par
By specialising our result to particular types of states, we can easily 
recover the entropic conditions for quantum error correction and the
conditions for saturation in the Holevo bound. In the general case, our
theorem characterises exactly when a quantum operation preserves correlations,
whether they be classical, in the form of pure entanglement, or more exotic,
such as combinations of the two or even bound 
entanglement~\cite{horodecki:bound}.
\par
We left open the problem of a similar characterisation in infinite dimension
(of system $B$ --- infinite $A$ and $C$ are covered by our result):
we only note that our method will certainly not work, as it relies ultimately 
on the classification of finite dimensional operator algebras.
A further interesting problem could be to address the \emph{approximate} 
case: if a state \emph{almost} satisfies strong subadditivity, does it 
mean that its structure is close in some sense to the form of 
theorem~\ref{thm:SSA:eq}? There might be a relation to~\cite{barnum:knill}
(see remark~\ref{rem:barnum-knill}), where an approximate fidelity condition
was studied.

\acknowledgments
We would like to thank Chris Fuchs, Mary Beth Ruskai and Ben Schumacher
for helpful discussions.
PH acknowledges the support of the Sherman Fairchild Foundation and 
the U.S. National Science Foundation through grant no. EIA-0086038.
DP was partially supported by Hungarian OTKA T032662.
The work of RJ and AW was supported by the U.K.~Engineering and Physical
Sciences Research Council.

\appendix

\section{An operator algebraic derivation of the Koashi--Imoto theorem}
\label{sec:operator:Koashi}
Let $\rho_1,\ldots,\rho_K$ be density operators on the finite--dimensional
Hilbert space ${\cal H}$. We are interested in the quantum operations
(completely positive, trace--perserving linear maps)
$T:{\cal B}({\cal H})\rightarrow{\cal B}({\cal H})$ which leave these
states invariant:
\begin{equation}
  \label{eq:invariance}
  \forall k\quad T\rho_k=\rho_k.
\end{equation}
By possibly shrinking ${\cal H}$ to the minimum joint supporting subspace
of the $\rho_k$, we may assume that the support of $\frac{1}{K}\sum_k \rho_k$ is
${\cal H}$, which we shall do in the following.
\par
>From the Stinespring dilation theorem~\cite{stinespring} it follows that
every such $T$ can be represented as
\begin{equation}
  \label{eq:stinespring}
  T\sigma = \tr_{\cal E} \bigl( U_{\cal HE} (\sigma\otimes\varepsilon) U_{\cal HE}^* \bigr),
\end{equation}
with another Hilbert space ${\cal E}$, a state $\varepsilon$ on
it, and a unitary acting on ${\cal H}\otimes{\cal E}$.
\par
In~\cite{koashi:imoto} the following result is proved by an explicit
algorithmic construction:
\begin{thm}[Koashi, Imoto~\cite{koashi:imoto}]
  \label{thm:koashi}
  Associated to the states $\rho_1,\ldots,\rho_K$ there exists
  a decomposition of ${\cal H}$ as
  \begin{equation}
    \label{eq:decomposition}
    {\cal H}=\bigoplus_j {\cal J}_j\otimes{\cal K}_j
  \end{equation}
  into a direct sum of tensor products, such that:\par\noindent
  $1$. The states $\rho_k$ decompose as
  $$\rho_k=\bigoplus_j q_{j|k}\rho_{j|k}\otimes\omega_j,$$
  where $\rho_{j|k}$ is a state on ${\cal J}_j$,
  $\omega_j$ is a state on ${\cal K}_j$ (which is independent
  of $k$), and $(q_{j|k})_j$ is a probability distribution over $j$'s.
  \par\noindent
  $2$. For every $T$ which leaves the $\rho_k$ invariant, every associated
  unitary from eq.~(\ref{eq:stinespring}) has the form
  $$U_{\cal HE}=\bigoplus_k \1_{{\cal J}_j}\otimes U_{{\cal K}_j{\cal E}},$$
  with unitaries $U_{{\cal K}_j{\cal E}}$ on
  ${\cal K}_j\otimes{\cal E}$ that satisfy
  $$\forall j\quad
     \tr_{\cal E}\bigl(U_{{\cal K}_j{\cal E}}
                           (\omega_j\otimes\varepsilon)
                       U_{{\cal K}_j{\cal E}}^*\bigr)
     =\omega_j.$$
\end{thm}
The purpose of this appendix is to present a short (but non--constructive)
proof of this theorem, based on the theory of operator algebras. Property
1~of the theorem has previously appeared in a paper of
Lindblad~\cite{lindblad:nocloning} and our approach closely follows the 
one taken there.
\par
We begin with a slight reformulation of the result, avoiding
the environment system ${\cal E}$:
\begin{prop}
  In the above theorem, property 2~is equivalent to\par\noindent
  $2'$. For every $T$ which leaves the $\rho_k$ invariant,
  $$\forall j\qquad T|_{{\cal B}({\cal J}_j\otimes{\cal K}_j)} = \id \otimes T_j,$$
  with $\id$ on ${\cal J}_j$ and $T_j$ on ${\cal K}_j$ such that
  $T_j(\omega_j)=\omega_j$.
\end{prop}
\begin{beweis}
  Clearly, $2$ implies $2'$. In the other direction, consider any
  $U$ implementing $T$. Clearly, because of $2'$,
  $$U|_{({\cal J}_j\otimes{\cal K}_j){\cal E}}
                 = \1_{{\cal J}_j}\otimes U_{{\cal K}_j{\cal E}},$$
  which yields the form $2$ for $U$.
\end{beweis}
\begin{beweis}[of theorem~\ref{thm:koashi}]
  Consider the set of quantum operations
  $${\bf F}=\bigl\{ F : \forall k\ F\rho_k=\rho_k \bigr\},$$
  which is obviously non--empty since it contains $T$ and $\id$.
  \par
  With each $F\in{\bf F}$ we associate the set
  $${\cal A}_F = \bigl\{ X\in{\cal B}({\cal H}) : F^*(X)=X \bigr\}$$
  of operators left invariant by the adjoint map $F^*$.
  By lemma~\ref{lemma:lindblad} this is a $*$--subalgebra
  of ${\cal B}({\cal H})$ and, in fact, if $F^*(X)=\sum_i B_i^*XB_i$,
  $${\cal A}_F=\{B_i,B_i^*\}'=\{X:\forall i\ XB_i=B_iX,\ XB_i^*=B_i^*X\}$$
  is the commutator of the Kraus operators of $F^*$.
  By the same lemma, this algebra furthermore is the image of
  ${\cal B}({\cal H})$ under the projection map
  $$P^* = \lim_{N\rightarrow\infty} \frac{1}{N}\sum_{n=1}^N (F^*)^n,$$
  whose adjoint is
  $$P = \lim_{N\rightarrow\infty} \frac{1}{N}\sum_{n=1}^N F^n.$$
  Clearly, $P\in{\bf F}$.
  Next, define
  $${\cal A}_0=\bigcap_{F\in{\bf F}} {\cal A}_F,$$
  which clearly is a $*$--subalgebra itself. Because all dimensions are finite,
  it can actually be presented as a finite intersection
  $${\cal A}_0={\cal A}_{F_1}\cap\ldots\cap{\cal A}_{F_M},$$
  and in fact there is $F_0\in{\bf F}$ such that ${\cal A}_0={\cal A}_{F_0}$.
  We may take, for example,
  $$F_0=\frac{1}{M}\sum_{\mu=1}^M F_\mu$$
  and use lemma~\ref{lemma:lindblad}. Denote the projection
  onto ${\cal A}_0$ derived from $F_0^*$ by $P_0^*$.
  \par
  Lemma~\ref{lemma:form:of:subalg} gives us the form of ${\cal A}_0$:
  $${\cal A}_0 = \bigoplus_j {\cal B}({\cal H}_{b^L_j})\otimes\1_{b^R_j},$$
  and, likewise, of $P_0^*$:
  $$P_0(\xi) = \bigoplus_j \tr_{b^R_j}\bigl(\Pi_j\xi\Pi_j\bigr) \otimes \omega_j.$$
  Thus, we obtain the advertised form of the states:
  $$\rho_k = P_0(\rho_k) = \bigoplus_j q_j\rho_{j|k}\otimes\omega_j.$$
  \par
  As for the properties of $T$, because ${\cal A}_T\supset{\cal A}_0$,
  we have $T^*|_{{\cal A}_0}=\id_{{\cal A}_0}$. More explicitly, for
  $A\in{\cal B}({\cal J}_j)$ and $\1\in{\cal B}({\cal K}_j)$,
  $$T^*(A\otimes\1) = A\otimes\1.$$
  Now assume $0\leq A\leq \1$, and consider $B\in{\cal B}({\cal K}_j)$
  such that $0\leq B\leq \1$. Then
  \begin{equation}
    \label{eq:sandwich}
    0 \leq T^*(A\otimes B) \leq T^*(A\otimes\1) = A\otimes\1 \leq \1\otimes\1.
  \end{equation}
  This implies that $T^*$ maps ${\cal B}({\cal J}_j\otimes{\cal K}_j)$
  into itself for all $j$, and hence the same applies to $T$.
  \par
  Now, eq.~(\ref{eq:sandwich}) applied with the rank--one projector
  $A=\ketbra{\psi}$, yields that
  $$T^*(\ketbra{\psi}\otimes B) = \ketbra{\psi}\otimes B',$$
  with $B'$ depending linearly on $B$. Dependence on $\ketbra{\psi}$ quickly
  leads to contradiction, so
  $$T^*(A\otimes B) = A\otimes T_j^*(B),$$
  which gives the desired form of $T$:
  $$T(\rho\otimes\sigma) = \rho\otimes T_j(\sigma),$$
  and application to the $\rho_k$ yields the invariance of the $\omega_j$
  under $T_j$.
\end{beweis}
\par\medskip
Here follow the general lemmas about unital completely positive maps
which were used in the proof of theorem \ref{thm:koashi}. The first one
is a mean ergodic theorem for the dual of a quantum operation. (The
statement is essentially the Kov\'acs-Sz\H ucs theorem ---
see e.g.~\cite{bratteli:robinson}, proposition 4.3.8 ---
but we give a proof in our setting.)
\begin{lemma}
  \label{lemma:lindblad}
  For a quantum operation $F$, the map
  $$P^*=\lim_{N\rightarrow\infty} \frac{1}{N}\sum_{n=1}^{N} (F^*)^n$$
  is a conditional expectation onto the $*$--subalgebra 
  $${\cal A}_F = \{ X : F^*(X)=X \} = \{ B_i,B_i^* \}'.$$
\end{lemma}
\begin{beweis}
  First of all, we want to see that ${\cal A}_F$ is a $*$--subalgebra.
  It is a linear subspace and the Kraus representation shows that if 
  $F^*(X)=X$, then $F^*(X^*)=\bigl(F^*(X)\bigr)^*=X^*$.
    \par
  With this, the Schwarz inequality (see e.g.~\cite{bratteli:robinson})
  gives that for invariant $X$,
  $$F^*(X^*X) \geq F^*(X^*)F^*(X)=X^*X.$$
  However, applying a faithful (i.e., non--degenenerate) invariant state,
  such as $\frac{1}{K}\sum_k \rho_k$, leaves only the possibility of equality:
  $$F^*(X^*X)=X^*X,$$
  from which it follows straightforwardly that the product of 
  invariant operators is again invariant.
  \par
  For $X\in {\cal A}_F$, one can confirm by direct calculation that
  $$\sum_i [X,B_i]^*[X,B_i] = F^*(X^*X)-X^*X = 0,$$
  the latter by the previous observation that $X^*X$ is also invariant.
  But since the left hand side is a sum of positive terms, all of them must be
  $0$, hence $[X,B_i]=0$ for all $i$. Similarly, $[X,B_i^*]=0$ for all $i$.
  \par
  These facts together say that ${\cal A}_F\subset\{ B_i,B_i^* \}'$, while the
  opposite containment is trivial.
  \par
  Another application of the Schwarz inequality gives that $F^*$ is a
  contraction.
  Hence the mean ergodic theorem for a contraction
  implies that the limit in the statement exists. (Due to the finite
  dimensional situation all the relevant topologies coincide.) To see
  $P^*(X) \in {\cal A}_F$, we compute
  $$(F^* P^*)(X) - P^*(X)  
      =  \lim_{N\rightarrow\infty} \frac{1}{N} \left( (F^*)^{N+1}(X) - F^*(X) \right)\!,$$ 
  which is clearly $0$ so the image of $P^*$ is contained in ${\cal A}_F$.
  Since $P^*(X)=X$ when $X \in {\cal A}_F$, it is also onto and a projection.
\end{beweis}
\par
\begin{lemma}
  \label{lemma:form:of:subalg}
  Let ${\cal A}$ be a $*$--subalgebra of ${\cal B}({\cal H})$, with a finite
  dimensional ${\cal H}$. Then there is a direct sum decomposition
  $${\cal H}=\bigoplus_j {\cal H}_{b^L_j}\otimes{\cal H}_{b^R_j},$$
  such that
  $${\cal A}=\bigoplus_j {\cal B}\bigl({\cal H}_{b^L_j}\bigr)\otimes\1_{b^R_j}.$$
  Any completely positive and unital projection $P^*$ of ${\cal B}({\cal H})$
  onto ${\cal A}$ is of the form
  $$P^*(X)
    = \bigoplus_j \tr_{b^R_j}\bigl( \Pi_jX\Pi_j(\1_{b^L_j}\otimes\omega_j) \bigr)
                                                                \otimes \1_{b^R_j},$$
  with the projections $\Pi_j$ onto the subspaces
  ${\cal H}_{b^L_j}\otimes{\cal H}_{b^R_j}$, and states $\omega_j$ on
  ${\cal H}_{b^R_j}$.
\end{lemma}
\begin{beweis}
  See~\cite{takesaki}, section I.11.
\end{beweis}




\begin{thebibliography}{99}
  \bibitem{AF} L. Accardi, A. Frigerio, ``Markovian cocycles'', 
  Proc.~Proc.~Roy.~Irish Acad., vol. 83A, no. 2, pp. 251--263, 1983.

  \bibitem{barnum:knill} H. Barnum, E. Knill, ``Reversing quantum dynamics with
    near--optimal quantum and classical fidelity'', J. Math. Phys., vol. 43, no. 5,
    pp. 2097--2106, 2002.

  \bibitem{schumacher:et-al} H. Barnum, M. A. Nielsen, B. Schumacher,
    ``Information transmission through a noisy
    quantum channel'', Phys. Rev. A, vol. 57, no. 6, pp. 4153--4175, 1998.

  \bibitem{bratteli:robinson} O. Bratteli, D. W. Robinson, \emph{Operator algebras
    and quantum statistical mechanics. 1. C${}^*$- and W${}^*$--algebras, symmetry groups,
    decomposition of states}, 2nd ed., Texts and Monographs in Physics,
    Springer Verlag, New York, 1987.

  \bibitem{holevo:bound} A. S. Holevo, ``Bounds for the quantity of information
    transmitted by a quantum channel'', Probl. Inf. Transm., vol. 9, no. 3,
    pp. 177--183, 1973.

  \bibitem{horodecki:bound} M. Horodecki, P. Horodecki, R. Horodecki,
    ``Mixed-state entanglement and distillation: Is there a `bound' 
    entanglement in nature?'', Phys. Rev. Lett. vol. 80, pp. 5239--5242, 1998.

  \bibitem{koashi:imoto} M. Koashi, N. Imoto, ``Operations that do not disturb partially
    known quantum states'', Phys. Rev. A, vol. 66, no. 2, 022318, 2002.

  \bibitem{kullback:leibler} S. Kullback, R. A. Leibler, ``On information and sufficiency'',
    Ann. Math. Statistics, 1951.

  \bibitem{lieb:ruskai:SSA} E. H. Lieb, M. B. Ruskai, ``Proof of the strong subadditivity
    of quantum-mechanical entropy'', J. Math. Phys., vol. 14, pp. 1938--1941, 1973.

  \bibitem{lindblad:mono} G. Lindblad, ``Completely positive maps and entropy inequalities'',
    Comm. Math. Phys., vol. 40, 147--151, 1975.

  \bibitem{lindblad:exchange} G. Lindblad, ``Quantum entropy and quantum
    measurements'', in: C. Bendjaballah, O. Hirota, S. Reynaud (eds.),
    \emph{Quantum Aspects of Optical Communications}, Lecture Notes in Physics,
    vol. 378, pp. 71--80, Springer Verlag, Berlin, 1991.

  \bibitem{lindblad:nocloning} G. Lindblad, ``A general no--cloning theorem'',
    Lett. Math. Phys, vol. 47, pp. 189--196, 1999.

  \bibitem{von:neumann:entropy} J. von Neumann, ``Thermodynamik quantenmechanischer
    Gesamtheiten'', Nachr. der Gesellschaft der Wiss. G\"ott., pp. 273--291, 1927.
    (See also J. von Neumann, \emph{Mathematical Foundations of Quantum
    Mechanics}, Princeton University Press, Princeton, NJ, 1996.)

  \bibitem{ohya:petz} M. Ohya, D. Petz, \emph{Quantum Entropy and Its Use},
    Springer Verlag: Texts and Monographs in Physics, Berlin Heidelberg, 1993.

  \bibitem{petz:sufficient} D. Petz, ``Sufficient subalgebras and the relative entropy of
    states of a von Neumann algebra'', Comm. Math. Phys., vol. 105, no. 1, pp. 123--131, 1986.
   ``Sufficiency of channels over von Neumann algebras'', Quart. J. Math. Oxford Ser. (2),
   vol. 39, no. 153, pp. 97--108, 1988.

  \bibitem{petz:RE-equality} D. Petz, ``Monotonicity of quantum relative entropy
    revisited'', Rev. Math. Phys., vol. 15, pp. 79--91, 2003.

  \bibitem{ruskai:SSA:eq} M. B. Ruskai, ``Inequalities for Quantum Entropy:
    A Review with Conditions for Equality'', J. Math. Phys.,
    vol. 43, pp. 4358--4375, 2002.

  \bibitem{schumacher:qec} B. Schumacher, M. A. Nielsen,
    ``Quantum data processing and error correction'', Phys. Rev. A.,
    vol. 54, pp. 2629--2635, 1996.

  \bibitem{shannon:info} C. E. Shannon, ``A mathematical theory of communication'',
    Bell Syst. Tech. Journal, vol. 27, pp. 379--423, 623--656, 1948.

  \bibitem{shor:ent-break} P. W. Shor, ``Additivity of the Classical Capacity of
    Entanglement--Breaking Quantum Channels'', J. Math. Phys., vol. 43, pp. 4334--4340, 2002.

  \bibitem{stinespring} W. F. Stinespring, ``Positive functions on
    C${}^*$--algebras'', Proc. Amer. Math. Soc., vol. 6, pp. 211--216, 1955.

  \bibitem{takesaki} M. Takesaki, \emph{Theory of Operator Algebras I},
    Springer--Verlag, New York--Heidelberg--Berlin, 1979.

  \bibitem{tucci} R. R. Tucci, ``Separability of Density Matrices and
    Conditional Information Transmission'', e--print {\tt quant-ph/0005119}, 2000.
    Based on ``Quantum Entanglement and Conditional Information Transmission'',
    e--print {\tt quant-ph/9909041}, 1999.

  \bibitem{uhlmann} A. Uhlmann, ``Relative entropy and the Wigner--Yanase--Dyson--Lieb
    concavity in an interpolation theory'', Comm. Math. Phys., vol. 54, no. 1,
    pp. 21--32, 1977.

  \bibitem{umegaki} H. Umegaki, ``Conditional expectation in an operator algebra IV.
    Entropy and information'', K\={o}dai Math. Sem. Rep., vol. 14, pp. 59--85, 1962.

  \bibitem{wehrl} A. Wehrl, ``General properties of entropy'',
    Rev. Modern Phys., vol. 50, no. 2, pp. 221--260, 1978.

\end{thebibliography}
\end{document}